\documentclass[sigconf]{acmart}

\copyrightyear{2024}
\acmYear{2024}
\setcopyright{acmlicensed}
\acmConference[CIKM '24] {Proceedings of the 33rd ACM International Conference on Information and Knowledge Management}{October 21--25, 2024}{Boise, ID, USA.}
\acmBooktitle{Proceedings of the 33rd ACM International Conference on Information and Knowledge Management (CIKM '24), October 21--25, 2024, Boise, ID, USA}
\acmISBN{979-8-4007-0436-9/24/10}
\acmDOI{10.1145/3627673.3680030}

\usepackage{amsfonts,amssymb}

\newcommand{\ignore}[1]{}
\usepackage{xspace}

\newcommand{\ourname}{TWIN-V2\xspace}

\newcommand{\mbf}[1]{\mathbf{#1}}

\usepackage{amsmath}
\usepackage{multirow}
\usepackage{multicol}
\usepackage{caption}
\usepackage{subcaption}
\usepackage{latexsym}
\usepackage{mflogo}
\usepackage{graphics}
\usepackage{tabularx}
\usepackage{threeparttable}
\usepackage{balance}
\usepackage{multirow}
\usepackage{makecell}
\usepackage{bbding}
\usepackage{enumitem}

\usepackage[linesnumbered,ruled,vlined]{algorithm2e}

\AtBeginDocument{%
  \providecommand\BibTeX{{%
    \normalfont B\kern-0.5em{\scshape i\kern-0.25em b}\kern-0.8em\TeX}}}

\begin{document}




\title{TWIN V2: Scaling Ultra-Long User Behavior Sequence Modeling for Enhanced CTR Prediction at Kuaishou}


\author{Zihua Si}
\affiliation{
  \institution{Renmin University of China}
  \city{Beijing}\country{China}
  }
\email{zihua_si@ruc.edu.cn}

\author{Lin Guan}
\affiliation{
  \institution{Kuaishou Technology Co., Ltd.}
  \city{Beijing}\country{China}
  }
\email{guanlin03@kuaishou.com}

\author{Zhongxiang Sun}
\affiliation{
  \institution{Renmin University of China}
  \city{Beijing}\country{China}
  }
\email{sunzhongxiang@ruc.edu.cn}

\author{Xiaoxue Zang}
\affiliation{%
  \institution{Kuaishou Technology Co., Ltd.}
  \city{Beijing}\country{China}
  }
\email{zangxiaoxue@kuaishou.com}

\author{Jing Lu}
\affiliation{%
  \institution{Kuaishou Technology Co., Ltd.}
  \city{Beijing}\country{China}
  }
\email{lvjing06@kuaishou.com}

\author{Yiqun Hui}
\affiliation{
  \institution{Kuaishou Technology Co., Ltd.}
  \city{Beijing}\country{China}
  }
\email{huiyiqun@kuaishou.com}

\author{Xingchao Cao}
\affiliation{
  \institution{Kuaishou Technology Co., Ltd.}
  \city{Beijing}\country{China}
  }
\email{caoxingchao@kuaishou.com}

\author{Zeyu Yang}
\affiliation{
  \institution{Kuaishou Technology Co., Ltd.}
  \city{Beijing}\country{China}
  }
\email{yangzeyu03@kuaishou.com}

\author{Yichen Zheng}
\affiliation{
  \institution{Kuaishou Technology Co., Ltd.}
  \city{Beijing}\country{China}
  }
\email{zhengyichen@kuaishou.com}

\author{Dewei Leng}
\affiliation{%
  \institution{Kuaishou Technology Co., Ltd.}
  \city{Beijing}\country{China}
  }
\email{lengdewei@kuaishou.com}

\author{Kai Zheng}
\affiliation{%
  \institution{Kuaishou Technology Co., Ltd.}
  \city{Beijing}\country{China}
  }
\email{zhengkai@kuaishou.com}

\author{Chenbin Zhang}
\affiliation{%
  \institution{Kuaishou Technology Co., Ltd.}
  \city{Beijing}\country{China}
  }
\email{aleczhang13@yeah.net}

\author{Yanan Niu}
\affiliation{%
  \institution{Kuaishou Technology Co., Ltd.}
  \city{Beijing}\country{China}
  }
\email{niuyanan@kuaishou.com}

\author{Yang Song}
\affiliation{%
  \institution{Kuaishou Technology Co., Ltd.}
  \city{Beijing}\country{China}
  }
\email{ys@sonyis.me}

\author{Kun Gai}
\affiliation{%
  \institution{Independent}
  \city{Beijing}\country{China}
  }
\email{gai.kun@qq.com}

\renewcommand{\shortauthors}{Zihua Si et al.}



\begin{abstract}
The significance of modeling long-term user interests for CTR prediction tasks in large-scale recommendation systems is progressively gaining attention among researchers and practitioners.
Existing work, such as SIM and TWIN, typically employs a two-stage approach to model long-term user behavior sequences for efficiency concerns. 
The first stage rapidly retrieves a subset of sequences related to the target item from a long sequence using a search-based mechanism namely the General Search Unit (GSU), while the second stage calculates the interest scores using the Exact Search Unit (ESU) on the retrieved results.
Given the extensive length of user behavior sequences spanning the entire life cycle, potentially reaching up to \(10^6\) in scale, there is currently no effective solution for fully modeling such expansive user interests. 
To overcome this issue, we introduced \ourname, an enhancement of TWIN, where a \emph{divide-and-conquer} approach is applied to compress life-cycle behaviors and uncover more accurate and diverse user interests.
Specifically, a hierarchical clustering method groups items with similar characteristics in life-cycle behaviors into a single cluster during the offline phase.
By limiting the size of clusters, we can compress behavior sequences well beyond the magnitude of $10^5$ to a length manageable for online inference in GSU retrieval.
Cluster-aware target attention extracts comprehensive and multi-faceted long-term interests of users, thereby making the final recommendation results more accurate and diverse.
Extensive offline experiments on a multi-billion-scale industrial dataset and online A/B tests have demonstrated the effectiveness of \ourname. 
Under an efficient deployment framework, \ourname has been successfully deployed to the primary traffic that serves hundreds of millions of daily active users at Kuaishou.
\end{abstract}



\maketitle

\section{Introduction}

Click-Through Rate (CTR) prediction is crucial for internet applications.
For instance, Kuaishou\footnote{\url{https://www.kuaishou.com/en}}, one of China's largest short-video sharing platforms, has employed CTR prediction as a core component of its ranking system.
Recently, much effort~\cite{DIN, SDIM, TWIN} has been devoted to modeling users' long-term historical behavior in CTR prediction.
Due to the extensive length of behaviors across the life cycle, modeling life-cycle user behaviors presents a challenging task.
Despite ongoing efforts to extend the length of historical behavior modeling in existing research, no method has yet been developed that can model a user's entire life cycle, encompassing up to 1 million behaviors within an app.

\begin{figure}[t]
  \centering
  \includegraphics[width=0.41\textwidth]{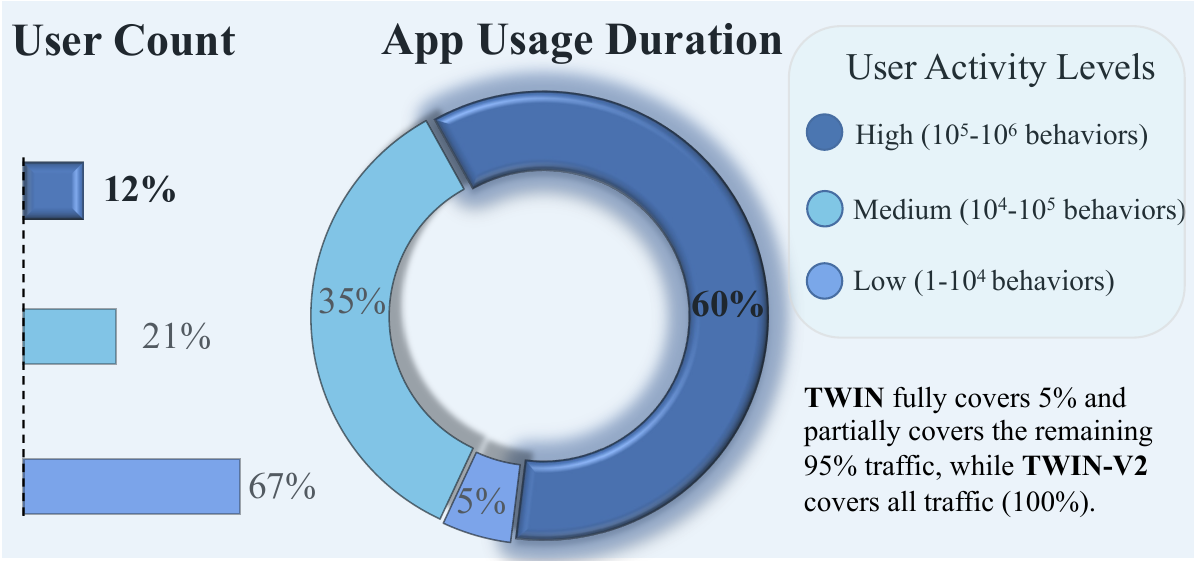}
  \vspace{-5pt}
  \caption{The Proportion of user count and app usage duration at Kuaishou. We grouped users based on the number of videos they have viewed over the past three years.
  Highly active users, though a small percentage (12\%), contribute the most usage duration (60\%).
  } 
  \vspace{-15pt}
  \label{fig: intro}
\end{figure}

Modern industrial systems~\cite{SIM, SDIM}, to utilize as long a user history as possible within a controllable inference time, have adopted a two-stage approach.
Specifically, in the first stage, a module called the General Search Unit (GSU) is used to filter long-term historical behaviors, selecting items related to the target item.
In the second stage, a module called the Exact Search Unit (ESU) further processes the filtered items, extracting the user's long-term interests through target attention.
The coarse-grained pre-filtering by the GSU enables the system to model longer historical behaviors at a faster speed during online inference.
Many approaches have tried different GSU structures to improve the accuracy of pre-filtering, such as ETA~\cite{ETA}, SDIM~\cite{SDIM} and TWIN~\cite{TWIN}.

Despite their effectiveness, existing first-stage GSUs generally have length limitations and cannot model user behaviors throughout the life cycle.
For instance, SIM, ETA, and SDIM can filter historical behaviors up to a maximum length of \(10^3\). TWIN extends the maximum length at the \(10^4\) to \(10^5\) level.
During deployment, TWIN utilizes the recent $10,000$ behaviors as inputs for GSU.
Unfortunately, these $10,000$ behaviors only cover the user's history for the last 3-4 months in the Kuaishou app, failing to encompass the entire life cycle of user behavior.
As illustrated in~\autoref{fig: intro}, we analyzed user behavior in the Kuaishou app over the past three years.
The medium and high user groups can view between \(10^4\) to \(10^6\) videos over three years, contributing the majority of the app's usage duration ($60\% + 35\% = 95\%$).
Hence, modeling the full life-cycle behaviors may enhance user experience and boost the platform's commercial gains.

To overcome this issue, we propose \ourname, an enhancement of TWIN, enabling it to model the full life cycle of user behavior.
\ourname employs a \emph{divide-and-conquer} approach, decomposing the full life-cycle history into different clusters and using these clusters to model the user's long-term interests.
Specifically, we divide the model into two parts: offline and online.
In the offline part, we employ hierarchical clustering to aggregate similar items in life-cycle behaviors into clusters and extract features from each cluster to form a virtual item.
In the online part, we utilize cluster-aware target attention to extract long-term interests based on clustered behaviors. Here, the attention scores are reweighted based on the corresponding cluster size.
Extensive experiments reveal that \ourname enhances performance across various types of users, leading to more accurate and diverse recommendation results. 

The main contributions are summarized as follows:

\noindent$\bullet$ We propose \ourname to capture user interests from extremely long user behavior, which successfully extends the maximum length for user modeling across the life cycle. 

\noindent$\bullet$ The hierarchical clustering method and cluster-aware target attention improve the efficiency of long sequence modeling and model more accurate and diverse user interests. 

\noindent$\bullet$ Extensive offline and online experiments have validated the effectiveness of \ourname in scaling ultra-long user behavior sequences.
    
\noindent$\bullet$ We share hands-on practices for deploying \ourname, including offline processing and online serving, which serve the main traffic of around 400 million active users daily at Kuaishou.

\section{Related Work}

\noindent\textbf{Click-Through Rate Prediction.}
Click-Through Rate (CTR) prediction is vital for modern Internet businesses.
Pioneering studies leveraged shallow models, like LR~\cite{LR}, FM~\cite{FM}, and FFM~\cite{FFM}, to model feature interactions. 
Wide\&Deep~\cite{WideDeep}, DeepFM~\cite{DeepFM}, and DCN~\cite{DCN,DCNv2} successfully combine deep and shallow models.
Furthermore, several studies have explored more complex neural networks, such as AutoInt~\cite{AutoInt} leveraging the multi-head self-attention mechanism.
Users' behavioral data have gained much attention recently.
YoutubeDNN~\cite{YoutubeDNN} has employed average pooling across the entire sequence of past behaviors.
Furthermore, DIN~\cite{DIN} leveraged a target attention mechanism to adaptively capture user interests from historical behaviors concerning the target item.
This target attention approach has been widely used by practitioners~\cite{DIN,DIEN,ETA,SDIM,TWIN}.
Following this line, DIEN~\cite{DIEN} and DSIN~\cite{DSIN} incorporated temporal information into the target attention.
BST~\cite{BST} and TWIN~\cite{TWIN} leveraged the multi-head attention for the target attention.

\begin{table}[t]
\footnotesize
\caption{Comparison with SOTA methods. The lower part lists the two-stage models. `Length' means the maximum sequence length of user behaviors in original papers.
}
\vspace{-10px}
\begin{tabular}{llc}
\toprule
\multirowcell{2}{Method} & \multirowcell{2}{Length} & \multirowcell{2}{GSU Stragegy}   \\
\\
\midrule
DIN \cite{DIN} & $\sim 10^3$ & N/A  \\
DIEN \cite{DIEN} & $\sim 10^2$ & N/A \\
MIMN \cite{MIMN} & $\sim 10^3$ & N/A  \\
DGIN \cite{DGIN_Meituan} & $\sim 10^4$ & N/A  \\
\midrule
UBR4CTR \cite{UBR4CTR1,UBR4CTR2} & $\sim 10^2$ & BM25  \\
SIM Hard \cite{SIM} & $\sim 10^3$ & Category Filtering\\
SIM Soft \cite{SIM} & $\sim 10^3$ & Inner Product  \\
{ETA \cite{ETA}} &{$\sim 10^3$} & LSH \& Hamming  \\
{SDIM \cite{SDIM}} &{$\sim 10^3$} &Hash Collision  \\
TWIN \cite{TWIN} & $\sim 10^5$ & Efficient Target Attention (TA) \\
\ourname (ours) & $\gg 10^5$ & Hierarchical Clustering \& TA  \\
\bottomrule
\label{tab:comparison}
\end{tabular}
\vspace{-5px}
\begin{tablenotes}
    \item[1] \textbf{\ourname}: It can handle behavioral histories far exceeding $10^5$ by adjusting the clustering hyper-parameters. The settings in this paper can process behaviors up to $10^6$, sufficiently covering the entire life cycle of users at Kuaishou.
\end{tablenotes}
\vspace{-15px}
\end{table}

\noindent\textbf{Long-Term User Behavior Modeling.}
Modeling lifelong historical behaviors is crucial in CTR prediction.
Early efforts involved memory networks to capture and remember user interests, such as HPMN~\cite{HPMN} and MIMN~\cite{MIMN}.
Several approaches attempt to directly shorten the length of user behaviors, such as UBCS~\cite{clustering_sampling_SIGIR22} which samples sub-sequences from the entire history and clusters all candidate items to accelerate the sampling process, and DGIN~\cite{DGIN_Meituan} employs item IDs to deduplicate history, thereby compressing redundant items.
Given the inconsistency between long-term and short-term interests, recent works commonly consider them separately.
SIM~\cite{SIM} and UBR4CTR~\cite{UBR4CTR1, UBR4CTR2} employ a two-stage approach, first retrieving a subset of behaviors related to the target item from long-term history, and then modeling long-term interests with retrieved behaviors and short-term interests with latest history. 
Following them, ETA~\cite{ETA}  explored the locality-sensitive hashing method and the Hamming distance to retrieve relevant behaviors.
SDIM~\cite{SDIM} directly gathers behavior items that share the same hash signature with the candidate item.
TWIN~\cite{TWIN} improves the retrieval accuracy by leveraging an identical target-behavior relevance metric for both stages.
TWIN~\cite{TWIN} also extends the maximum length of lifelong behaviors around $10^3$ to $10^4-10^5$ by accelerating the target attention mechanism.

This paper aims to extend long-term history modeling to the life-cycle level while maintaining efficiency.
\autoref{tab:comparison} summarizes the comparison between \ourname and previous works.

\section{Method}

This section elaborates on the proposed model, detailing the entire process from the overall workflow to the deployment framework. 
Notations are summarized in Appendix~\ref{appendix: notation}.

\subsection{Preliminaries}

The core task of CTR prediction is to forecast the probability of a user clicking on an item.
Let $\mathbf{x}_k\in\mathbb R^{d_0}$ denote the feature representation of $k$-th data sample and let $y_k \in \{0,1\}$ denote the label of the $k$-th interaction.
The process of CTR prediction can be written as:
\begin{equation}
\hat y_k = \sigma(f(\mathbf x_k)),
\end{equation}
where $\sigma(\cdot)$ is the sigmoid function, $f$ is the mapping function implemented as the CTR model $f:\mathbb R^{d_0} \rightarrow \mathbb R$, and $\hat y_k$ is the predicted probability. The model $f$ is trained by the binary cross entropy loss:
\begin{equation}
    \mathcal{L} = -\frac{1}{|\mathcal D|} \sum_{k=1}^{|\mathcal D|} y_k \log(\hat y_k) + (1-y_k)\log(1-\hat y_k),
\end{equation}
where $\mathcal D$ denotes the training dataset.


\subsection{Overall Workflow}

\begin{figure*}[t]
  \centering
  \includegraphics[width=0.78\textwidth]{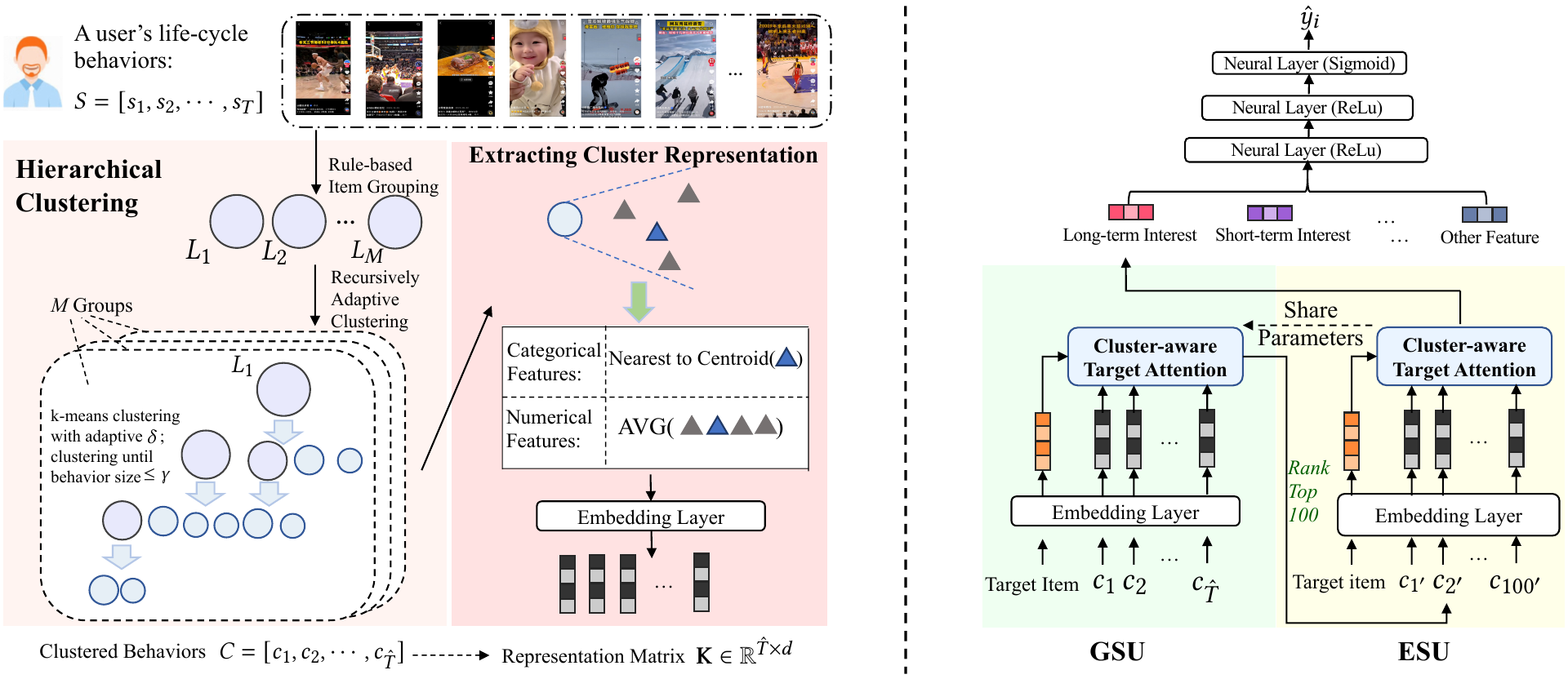}
  \vspace{-7pt}
  \caption{Overview of \ourname in Kuaishou's CTR prediction system.
  Left: the offline component compresses life-cycle behaviors and extracts features of clustered behaviors.
  Right: the online component captures user interests from life-cycle behaviors, which adopts a two-stage approach. GSU and ESU for both stages take clustered behaviors as input.
  } 
  \vspace{-10pt}
  \label{fig: model}
\end{figure*}

\autoref{fig: model} illustrates the overall framework of \ourname, including \emph{offline} and \emph{online} parts. 
This paper focuses on life-cycle behavior modeling. Thus we emphasize this part and omit other parts.

Since the behavior over the entire life cycle is too lengthy for online inferring, we initially compress it during the \emph{offline} phase.
User behavior often includes many similar videos, as users frequently browse their favorite topics.
Therefore, we employ hierarchical clustering to aggregate items with similar interests in each user's history into clusters. Online inferring captures the user's long-term interests using these clusters and their representation vectors.

During \emph{online inferring}, we adopt a two-stage approach for life-cycle modeling. First, we use GSU to retrieve the top 100 clusters related to the target item from the clustered behaviors, and then employ ESU to extract long-term interests from these clusters.
In GSU and ESU, we implement cluster-aware target attention, adjusting weights for different clusters.

\subsection{Life-Cycle User Modeling}
\label{Sec: method}

Considering that the user histories over the life cycle have ultra-long sequences, we first compress these behaviors in a divide-and-conquer method and then extract users' long-term interests from compressed behaviors.

\subsubsection{Hierarchical Clustering Over Life Cycle}
\label{sec: clustering}

\begin{algorithm}[t]
\footnotesize
\caption{Hierarchical Clustering Over Life Cycle}
\label{Algo: kmeans}
\KwIn{
History items $S_{1:T}$, Item embeddings $\mathbf{K}_{s}$, Item Playing Completion Ratio $P_{1:T}$, group number $M$, and maximum cluster size $\gamma$, where $s_j$ is the $j$-th item, $\mathbf{K}_s\in \mathbb{R}^{T\times d}$, $p_i \in \mathbb{R}$, and $M, \gamma \in \mathbb{N}^+$
}
\KwOut{Clustered items $C_{1:\hat{T}}$}
\SetKwFunction{}
\KwFunction{
\textbf{Function} $h_{\text{comp}(S_{1:T})}$ :
}
\SetAlgoLined
\BlankLine

\textcolor[rgb]{0.2,0.4,0.6}{\# Rule-based item grouping in lines 3-8}\\

Initialize $M$ empty lists \( L_1, L_2, \cdots, L_M \)\;
  Define a function \( \text{getGroup}(p) \) that returns the group number \( m \) for the $j$-th item considering its $p_j$;

\For{\(j = 1\) \KwTo \(T\)}{
    \( m = \text{getGroup}(p_j) \)\;
    \( L_m.\text{Append}(s_j) \)\;
  }

\textcolor[rgb]{0.2,0.4,0.6}{\# Recursively adaptive clustering in lines 10-28}\\

Initialize an empty list $C$;

\For{\(m = 1\) \KwTo \(M\)}{
    Initialize an empty queue $Q$\;
    $Q.\text{enqueue}(L_m)$\;
    \While{$Q$ is not empty}{
    $V$ \(\leftarrow\) $Q.\text{dequeue}()$ \;
        \If{$|V|<\gamma$}
        {$C.\text{append}(V)$}
        \Else{
        $\delta$ \(\leftarrow\) $\lfloor |V|^{0.3} \rfloor$\;
        \textcolor[rgb]{0.2,0.4,0.6}{\# Cluster items in $V$ into $\delta$ groups based on their corresponding embeddings in $\mathbf{K}_s$}\\
        $J_{1:\delta}$\(\leftarrow\)Kmeans($\delta, V, \mathbf{K}_s$)\;
        \For{\(i = 1\) \KwTo $\delta$ }{
        $Q$.enqueue($J_i$)\;
        }
        }
    }
} 
\KwRet{\( C \)}\
\end{algorithm}

For a user \(u\), we represent all her historical behaviors as a sequence of items $S = [s_1,s_2,\cdots, s_T]$, where $T$ is the number of life-cycle behaviors and $s_j$ is the $j$-th interacted item.
Users may watch many similar videos in $S$. 
For instance, a user fond of NBA games may have hundreds of basketball-related videos in his history.
Therefore, a natural idea is to aggregate similar items in $S$ into clusters, using a single cluster to represent many similar items, thereby reducing the size of $T$.

Formally, we aim to use a compression function $h_{\text{comp}}$ to reduce a behavior sequence $S$ of length \(T\) to a sequence $C$ of length \(\hat T\):
\begin{equation}
    C = h_{\text{comp}}(S),
\end{equation}
where $\hat{T} \ll T$. Let $C$ denote the clustered behaviors $C=[c_1,c_2,\cdots,c_{\hat T}]$, where $c_i$ denotes the $i$-th set containing clustered items. 

Specifically, we employ a hierarchical clustering approach to implement \(h_{\text{comp}}\), as illustrated in Algorithm~\ref{Algo: kmeans}. 
In the first step, we simply split the historical behaviors into groups.
We first categorize historical items into $M$ different groups based on the proportion of time they were played by the user $u$, which is denoted as $p_j=\frac{\text{playing time}}{\text{video duration}}$ for the $j$-th behavior item $s_j$. 
In the second step, we recursively cluster the life-cycle historical behaviors of each group until the number of items in each cluster does not exceed \(\gamma\).
We employ the widely used k-means method with adaptive cluster size $\gamma$ to cluster the embeddings $\mathbf{K}_s$ of behaviors $S$ obtained from the recommendation model.  $\gamma$ is a hyper-parameter to control the maximum cluster size.

Next, we will elaborate on the rationales behind Algorithm~\ref{Algo: kmeans}:

\noindent$\bullet$ \textbf{Item Grouping:} 
In short video scenarios, the playing completion ratio of a video can be seen as an indicator of the user's level of interest in the video.
We initially group behaviors by playing completion ratio to ensure the final clustering results have a relatively consistent playing time ratio.
If this is not done, it would result in an imbalanced distribution within each cluster because k-means only considers item embeddings.
The function $\text{getGroup}(p_j)$ is specific to the application scenarios, for example, dividing the proportional range into five equal parts.
$M$ is a constant, set to 5 in our practices.

\noindent$\bullet$ \textbf{Recursively Adaptive Clustering}: We opt for hierarchical clustering instead of single-step clustering because user history is personalized, making it impractical to set a universal number of clustering iterations.
We also dynamically set the number of clusters (\( \delta \) equals the $0.3$ power of the number of items), allowing clustering to adapt to varying scales of behaviors.
The clustering process stops when the item size falls below \( \gamma \), where $\gamma$ is set to 20.
In our practice, the average cluster size is approximately \( \frac{\gamma}{2} \), hence \( \hat{T} \) is about $\frac{T}{10}$, significantly reducing the length of life-cycle behaviors.
Item embeddings $\mathbf{K}_s$ used for clustering are obtained from the recommendation model, which means the clustering is guided by collaborative filtering signals.

Thus, we compress the originally lengthy user behavior sequence $S = [s_1,s_2,\cdots, s_T]$ into a 
shorter sequence $C=[c_1,c_2,\cdots,c_{\hat T}]$ by aggregating similar items into clusters.

\subsubsection{Extracting Cluster Representation}
After obtaining clusters $C$, we need to extract features from items of each cluster. 
To minimize computational and storage overhead, we use a virtual item to represent the features of each cluster.

We divide item features into two categories (numerical and categorical) and extract them using different methods.
Numerical features are typically represented using scalers, for example, video duration and user playing time for a video.
Categorical features are commonly represented using one-hot (or multi-hot) vectors, such as video ID and author ID.
Formally, given an arbitrary item $v$, its feature can be written as:
\begin{equation}
\small
\label{instance}
    \mathbf{x}_v = [\underbrace{\mathbf{x}_{c,1}^{(v)};\mathbf{x}_{c,2}^{(v)};\dots;\mathbf{x}_{c,N_1}^{(v)}}_{\text{categorical (one-hot)}};\underbrace{x_{s,1}^{(v)};x_{s,2}^{(v)};\dots;x_{s,N_2}^{(v)}}_{\text{numerical (scaler)}}],
\end{equation}
For simplicity, we use $\mathbf{x}^{(v)}_{1:N_1}$ to denote categorical features and $\mathbf{x}^{(v)}_{1:N_2}$ to denote numerical features.
For a cluster $c_i$ in $C$, we calculate the average of all numerical features of the items it contains as $c_i$'s numerical feature representation:
\begin{equation}
    \mathbf{c}^{(i)}_{1:N_2} = \frac{1}{|c_i|}\sum_{v\in c_i} \mathbf{x}^{(v)}_{1:N_2},
\end{equation}
For categorical features, since their average is meaningless, we adopt the closest item to the centroid in cluster $c_i$ to represent cluster $c_i$:
\begin{gather}
    \mathbf{c}^{(i)}_{1:N_1} = \mathbf{x}^{(v)}_{1:N_1},\\
    v = \operatorname*{arg\,min}_{v \in c_i}
    \left\| \mathbf{k}_v - \mathbf{k}_{\textrm{centroid}} \right\|_2^2,
\end{gather}
where $\mathbf{k}_v, \mathbf{k}_{\textrm{centroid}}\in \mathbb{R}^d$ are embeddings of the item $v$ and centroid in $c_i$, respectively. These embeddings can be looked up from $\mathbf{K}_s$.
Thus, the entire cluster $c_i$ is represented by an aggregated, virtual item feature $\mathbf{c}_i = [\mathbf{c}^{(i)}_{1:N_1}; \mathbf{c}^{(i)}_{1:N_2}]$.
After passing through embedding layers, each cluster's feature is embedded into a vector, e.g., $\mathbf{k}_i \in \mathbb{R}^d$ for $\mathbf{c}_i$.
The GSU and ESU modules estimate the relevance between each cluster and the target item based on these embedded vectors.

\subsubsection{Cluster-aware Target Attention}

Following TWIN~\cite{TWIN}, we employ an identical efficient attention mechanism in both ESU and GSU, which take representations of clusters $C$ as input.

Given the clustered behaviors $C=[c_1,c_2,\cdots,c_{\hat T}]$, we obtain a matrix $\mathbf{K}\in \mathbb{R}^{\hat T \times d}$ composed of representation vectors through the embedding layer, where $\mathbf{k}_i \in \mathbb{R}^d$ is the vector for $i$-th cluster's feature $\mathbf{c}_i$.
Then we use a target attention mechanism to measure the relevance between the target item and historical behaviors.
We initially apply the "Behavior Feature Splits and Linear Projection" technique from TWIN~\cite{TWIN} to enhance the efficiency of target attention, details in Appendix~\ref{app: Behavior}. 
This method splits item embeddings into inherent and cross parts.
$\mathbf{q} \in \mathbb{R}^{H}$ denotes the vector of the target item's inherent features. 
$\mathbf{K}_h \in \mathbb{R}^{\hat T \times H}, \mathbf{K}_c \in \mathbb{R}^{\hat T \times C}$ are inherent and cross embeddings of clustered behaviors respectively, where $\mathbf{K} = [\mathbf{K}_h, \mathbf{K}_c]$.
We can calculate the relevance scores $\boldsymbol{\alpha}\in \mathbb{R}^{\hat T}$ between the target item and clustered behaviors as follows:
\begin{equation}\label{equ:alpha}
\boldsymbol{\alpha} = \frac{(\mathbf K_h \mathbf W^{h})(\mathbf q^\top \mathbf W^{q})^\top}{\sqrt d_k}+ (\mathbf K_c  \mathbf W^c)\boldsymbol{\beta},
\end{equation}
where $\mathbf W^{h}, \mathbf W^{q}, \mathbf W^{c}$ are linear projections, $d_k$ is the projected dimension, and $\boldsymbol{\beta}$ is a learnable cross feature weight. 
The $\boldsymbol{\alpha}$ inadequately represents the relationship between clusters and the target item, due to varying item counts in clusters. 
Assuming two clusters have the same relevance,  the cluster with more items should be considered more important, as more items imply a stronger user preference.
Thus, we adjust the relevance scores according to the cluster size:
\begin{equation}
\label{eq: alpha'}
    \boldsymbol{\alpha}' = \boldsymbol{\alpha} + \ln{\mathbf{n}},
\end{equation}
where $\mathbf{n} \in \mathbb{N}^{\hat T}$ denote the size of all clusters in $C$. Each element $n_i$ means the cluster size of $c_i$.

During the GSU stage, we use $\boldsymbol{\alpha}'$ to select the top $100$ clusters with the highest relevance scores from the clustered behaviors of length $\hat T$.
Then, these 100 clusters are input into the ESU, where they are aggregated based on relevance scores, resulting in a representation of the user's long-term interests:
\begin{equation}
\textrm{Attention}(\mathbf q^\top \mathbf W^q, \mathbf K_h \mathbf W^h, \mathbf K_c \mathbf W^c, \mathbf K \mathbf W^v)= \text{Softmax}(\boldsymbol \alpha')^\top \mathbf K \mathbf W^v,\\
\end{equation}
where $\textbf W^v$ is a projection matrix and the notations in this equation are slightly abused for simplicity by setting $\hat T=100$ for the ESU stage.
Please note that the $\text{Softmax}(\boldsymbol \alpha')$ can be written as:
\begin{equation*}
    \text{Softmax}(\boldsymbol \alpha')= \sum_{i=1}^{\hat T}\frac{n_i \cdot \exp{\alpha_i}}{\Sigma_{j=1}^{\hat T}n_j \cdot \exp{\alpha_j}},
\end{equation*}
where the relevance score of each cluster $\alpha_i$ is reweighted by cluster size $n_i$. We employ multi-head attention with $4$ heads to obtain the ultimate long-term interest:
\begin{equation}
\begin{aligned}
&\text{Long-term Interests} = \text{Concat}(\text{head}_1,...,\text{head}_4)\mathbf W^o, \\
\text{head}_a &= \text{Attention}(\mathbf q^\top \mathbf W_a^q, \mathbf K_h \mathbf W_a^h, \mathbf K_c \mathbf W_a^c, \mathbf K \mathbf W_a^v), a\in\{1,...,4\},
\end{aligned}
\end{equation}
where $\mathbf W^o$ is a projection. 

\textbf{Remark.} Utilizing behaviors derived from hierarchical clustering in GSU enables the model to accommodate longer historical behaviors, achieving life-cycle level modeling in our systems.
Compared to TWIN, GSU's input of life-cycle behaviors encompasses a broader range of historical interests, thereby offering a more comprehensive and accurate modeling of user interests.
Additionally, while existing works use the top 100 retrieved behaviors as input for ESU, \ourname utilizes 100 clusters. 
These clusters cover more behaviors beyond 100, allowing ESU to model a broader spectrum of behaviors.
All these advantages lead to more accurate and diverse recommendation results, verified in Section~\ref{sec: online AB}.


\subsection{Deployment of \ourname}
\label{sec: deployment}

\begin{figure}[t]
  \centering
  \includegraphics[width=0.45\textwidth]{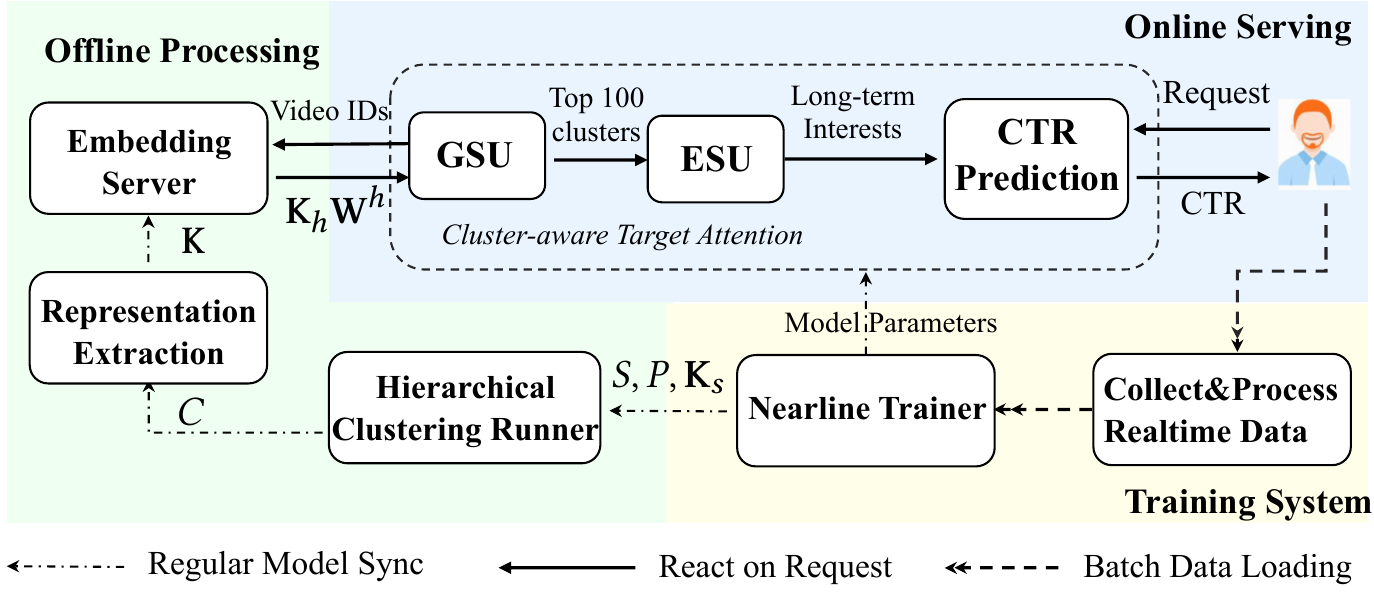}
  \vspace{-5pt}
  \caption{
  The deployment of \ourname at Kuaishou.
  } 
  \label{fig: deployment}
  \vspace{-15px}
\end{figure}

We divide our hands-on practices for deploying \ourname into two parts: online and offline, as shown in~\autoref{fig: deployment}.

\textbf{Overview. }
In~\autoref{fig: deployment}, we illustrate with a single user example. 
When the system receives a user's request, it first extracts behavioral features from the offline-processed user life-cycle behaviors. Then, using GSU and ESU, it models these into long-term interest inputs for the CTR model to make predictions.
We employ a nearline trainer for real-time model training, which incrementally updates model parameters using real-time user interaction data within an 8-minute window.
For life-cycle behaviors, we apply hierarchical clustering and feature extraction to process them. This is conducted through periodic offline processing and updates.

\textbf{Offline Processing.}
Offline processing aims to compress the user's entire life-cycle behaviors.
Considering Kuaishou's user count is on a billion scale, we periodically compress their life-cycle behaviors.
The hierarchical clustering runner performs a full update every 2 weeks.
We set the maximum cluster size $\gamma$ in hierarchical clustering to $20$, leading to an average cluster size around $10$.
Through cluster representation extraction, each cluster is aggregated into a virtual item.
In our practices, historical behavior is compressed to $10\%$ of its original length, resulting in a $90\%$ reduction in storage costs.
We employ the inherent feature projector proposed in TWIN for the embedding server, refreshing its parameters every 15 minutes from the latest CTR model.
 
\textbf{Online Serving. }
After the user sends a request to the system, the offline system sends data features to the GSU, which calculates the behavior-target relevance scores, $\boldsymbol{\alpha}'$, according to Eq.~\eqref{equ:alpha}~and~\eqref{eq: alpha'}.
Then, the ESU selects and aggregates the top 100 clustered behaviors as long-term interests for the CTR model, which makes a prediction.
To ensure efficiency in online inference, we maintain the precomputing and caching strategies from TWIN.

\ourname was incrementally deployed with TWIN. TWIN models the behaviors in recent months, while TWIN-V2 models the behaviors throughout the entire life cycle, allowing them to complement each other.

\section{Experiment}

In this section, we verify the effectiveness of \ourname by conducting extensive offline and online experiments.

\subsection{Experimental Setup}

\subsubsection{Dataset}
Since \ourname is designed for user behaviors with extremely long lengths, it's necessary to use datasets with ample user history.
To the best of our knowledge, there is no existing public dataset with an average history length exceeding \(10^4\).
Therefore, we extracted user interaction data from the Kuaishou app over five consecutive days to serve as the training and testing sets.
Samples were constructed from clicking logs with click-through as labels. 
To capture users' life-cycle histories, we further retraced each user's past behavior, covering the maximum history length for each user at $100,000$.
\autoref{tab:dataset} shows the basic statistics of this industrial dataset.
The first 23 hours of each day were used as the training set, while the final hour served as the testing set.

\begin{table}[t]
\footnotesize
\caption{Statistics of the large-scale industrial dataset collected from Kuaishou.}
\vspace{-5px}
\begin{tabular}{llc}
\toprule
Data & Field & Size \\ 
\midrule
\multirow{4}{*}{Daily Log Info} & Users & 345.5 million \\
& Videos & 45.1 million \\
& Samples & 46.2 billion \\
& Average User Actions & 133.7 / day \\
\midrule
\multirow{2}{*}{Historical Behaviors} & Average User Behaviors & 14.5 thousand \\
& Max User Behaviors & 100 thousand \\
\bottomrule
\label{tab:dataset}
\end{tabular}
\vspace{-20px}
\end{table}

\subsubsection{Baselines}

Following common practices~\cite{SDIM, TWIN}, given that our focus is on modeling extremely long user histories, we compare \ourname with the following SOTA baselines:

(1) \textbf{Avg-Pooling} : A naive method leverages average pooling. (2) \textbf{DIN}~\cite{DIN}: It introduces the target attention for recommendation. (3) \textbf{SIM Hard}~\cite{SIM}: Hard-search refers to the process where GSU filters long-term history based on its category. (4) \textbf{SIM Soft}~\cite{SIM}: Soft search involves selecting the top-k items by computing the vector dot product between the target item and items in the history. (5) \textbf{ETA}~\cite{ETA}: It employs Locality Sensitive Hashing to facilitate end-to-end training. (6) \textbf{SDIM}~\cite{SDIM}: The GSU aggregates long-term history by gathering behaviors that share the same hashing signature with the target item. (7) \textbf{SIM Cluster}:  Due to the reliance on annotated categories in SIM Hard, we improved it by the clustering of video embeddings. GSU retrieves history behaviors within the same cluster with the target item. Here we clustered all items into $1,000$ groups. (8) 
\textbf{SIM Cluster+}: This is an advanced variant of SIM Cluster with increasing cluster number to $10,000$. (9) \textbf{TWIN}~\cite{TWIN}: It proposes an efficient target attention for GSU and ESU stages.


To ensure a fair comparison, we maintained consistency in all aspects of the models except for the long-term interest modeling. 

\subsubsection{Evaluation Metrics \& Protocal}
We split the dataset into training and testing sets based on timestamps. 
We utilized data from 23 consecutive hours each day for the training set and the remaining 1 hour's data for the test set.
For all models, we evaluated their performance over five consecutive days and reported the average results.
As for evaluation metrics, we adopted \textbf{AUC} (Area under the ROC curve) and \textbf{GAUC} (Group AUC), which are widely used.
We calculated the \textit{mean} and standard deviation (\textit{std}) of both metrics over 5 consecutive days.

\subsubsection{Implementation Details}

\begin{table}[t]
\small
\caption{
Overall comparison.
The best and the second-best performances are denoted in bold and underlined fonts, respectively.
An 0.001-level improvement is considered significant for CTR prediction, as supported by~\cite{DCNv2, ESMM_alibaba_SIGIR18}.
}
\vspace{-10px}
\begin{tabular}{lcc}
\toprule
\textbf{Method} & \textbf{AUC} (\textit{mean} \footnotesize{$\pm$ \textit{std}} ) $\uparrow$  & \textbf{GAUC} (\textit{mean} \footnotesize{$\pm$ \textit{std}} ) $\uparrow$  \\ 
\midrule
Avg-Pooling  & 0.7855 \footnotesize{$\pm$ 0.00023} & 0.7168 \footnotesize{$\pm$ 0.00019} \\
DIN  & 0.7873 \footnotesize{$\pm$ 0.00014} & 0.7191 \footnotesize{$\pm$ 0.00012} \\
\hline
SIM Hard  & 0.7901 \footnotesize{$\pm$ 0.00016} & 0.7224 \footnotesize{$\pm$ 0.00021} \\
ETA   & 0.7910 \footnotesize{$\pm$ 0.00004} & 0.7243 \footnotesize{$\pm$ 0.00011} \\
SIM Cluster  & 0.7915 \footnotesize{$\pm$ 0.00017} & 0.7253 \footnotesize{$\pm$ 0.00018} \\
SDIM  & 0.7919 \footnotesize{$\pm$ 0.00009} & 0.7267 \footnotesize{$\pm$ 0.00006} \\
SIM Cluster+  & 0.7927 \footnotesize{$\pm$ 0.00009} & 0.7275 \footnotesize{$\pm$ 0.00011} \\
SIM Soft  & 0.7939 \footnotesize{$\pm$ 0.00014} & 0.7299 \footnotesize{$\pm$ 0.00013} \\
TWIN  & \underline{0.7962 \footnotesize{$\pm$ 0.00008}} & \underline{0.7336 \footnotesize{$\pm$ 0.00011}} \\
\midrule
\ourname  & \textbf{0.7975 \footnotesize{$\pm$ 0.00010}} & \textbf{0.7360 \footnotesize{$\pm$ 0.00009}} \\
Improv.  & +0.16\% & +0.33\% \\
\bottomrule
\label{tab:overall}
\end{tabular}
\vspace{-30px}
\end{table}

For all models, we utilize the same item and user features in our industrial context, including ample attributes such as video ID, author ID and user ID.
\ourname limits the length of individual user history to a maximum of $100,000$ items for clustering life-cycle user behavior.
As discussed in Section~\ref{sec: deployment}, \ourname empirically compresses historical behavior to about $10\%$ of its original size, resulting in a maximum input length of approximately $10,000$ in the GSU.
For other two-stage models, the maximum length of historical behavior input into the GSU is limited to $10,000$ behaviors. 
For DIN and Avg-Pooling, the maximum length of recent history is $100$ since they are not designed for long-term behaviors.
All the two-stage models leverage the GSU to retrieve the top $100$ historical items and feed them into the ESU for interest modeling.
The batch size is set to 8192.
The learning rate is 5e-6 for Adam.

\subsection{Overall Performance}

From the results in~\autoref{tab:overall}, we have the following observations:

\noindent$\bullet$\textbf{\ourname significantly outperforms other baselines by a large margin.}
As supported by existing literature~\cite{DCNv2, ESMM_alibaba_SIGIR18}, an improvement of $0.001$ in AUC is considered significant for CTR prediction and is sufficient to yield online benefits.
\ourname achieves an improvement over the second-best performing TWIN model, with an increase of $0.0013$ in AUC and $0.0024$ in GAUC.
These improvements verify the effectiveness of \ourname.

\noindent$\bullet$\textbf{\ourname exhibited a higher relative improvement in GAUC compared to its relative improvement in AUC.}
GAUC provides a more fine-grained measure of models.
The larger improvement in GAUC indicates that \ourname achieves enhancements across various types of users, rather than just performing better on the overall samples, and it also excels in specific subsets.
Since \ourname incorporates longer behavioral sequences, we believe it achieves greater improvements among highly active user groups, a hypothesis validated in Appendix~\ref{sec: study on user groups}.

\subsection{Ablation Study}

We conducted an ablation study to investigate the role of core modules in \ourname and the rationale behind our design.

\subsubsection{Comparison of Different Hierarchical Clustering Methods}

Our hierarchical clustering method has two key features: 1. It utilizes a dynamic clustering number \(\delta\) to adapt to varying sizes of behaviors; 2. The k-means clustering is non-uniform, resulting in clusters of different sizes in the end.
We first verify the effectiveness of adaptive \(\delta\), creating a variant that uses a fixed cluster number $\delta=2$ for all cases, denoted as `Binary'.
Furthermore, to test the effect of uniform cluster sizes, we create a variant that enforces balanced k-means clustering results when \(\delta=2\), denoted as `Balanced\&Binary'. 
In this variant, after each k-means iteration, a portion of items from the larger cluster, which are closer to the centroid of the other cluster, are moved to the other cluster to ensure equal sizes of the final two clusters.

\begin{table}[t]
\footnotesize
\centering
\caption{ Statistical analyses for different hierarchical clustering methods.
`Adaptive' denotes our method. `Binary' denotes the variant with $\delta=2$. `Balanced\&Binary' denotes the variant with $\delta=2$ and enforced balanced k-means.
}
\vspace{-10px}
\begin{tabular}{c|ccc}
\toprule
 Method  & Cluster number $\delta$   & Cluster Accuracy &Running Time (sec) \\ \midrule
Balanced\&Binary            & 2                           & 0.765       &       1.010   \\ 
Binary & 2    &  0.789  & 1.201 \\
Adaptive (ours) &  dynamic   & 0.802   & 0.750 \\
\bottomrule
\end{tabular}
\begin{tablenotes}
    \item[1] Cluster accuracy is measured by the average cosine similarity of each item to the centroid within each cluster. 
    Running time denotes the average time taken for each user to complete hierarchical clustering.
\end{tablenotes}
\label{tab: different HC}
\vspace{-15px}
\end{table}

\begin{figure}[t]
    \centering
    \begin{subfigure}{0.4\linewidth}
        \includegraphics[width=\textwidth]{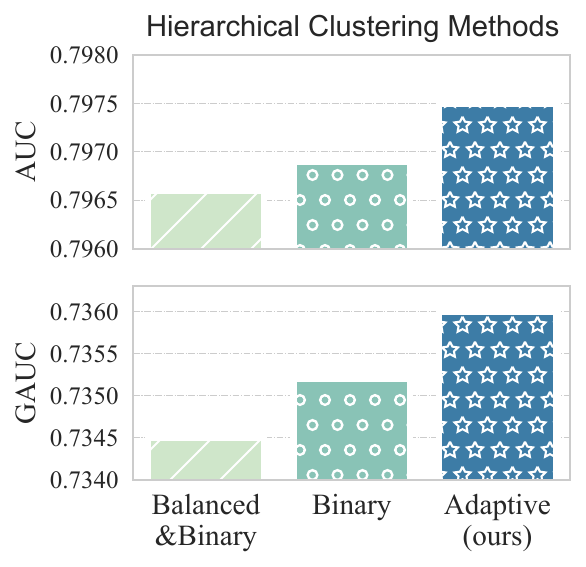}
    \end{subfigure}
    \hspace{1mm}
    \begin{subfigure}{0.4\linewidth}
        \includegraphics[width=\textwidth]{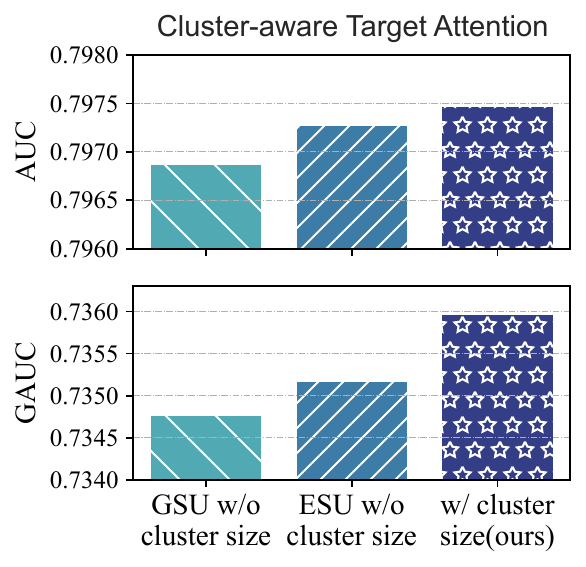}
    \end{subfigure}
    \vspace{-12px}
\caption{Effects of core modules in~\ourname.
Left: Comparisons between different hierarchical clustering methods.
Right: Impact of omitting cluster size in the reweighting of relevance scores.
}
  \label{fig: ablation study}
\vspace{-10px}
\end{figure}

\autoref{tab: different HC} reports the statistical analyses on \ourname and these two variants.
Among the three methods, ours achieves the highest cluster accuracy.
This suggests that our method creates clusters where items have more closely matched representation vectors, resulting in higher similarity among items within each cluster.
Our method also achieves the shortest running time, confirming the efficiency of adaptive $\delta$.
Furthermore, we reported the performance of these methods on the test set, as depicted in the left part of~\autoref{fig: ablation study}.
These results validate that the adaptive method can assign more similar items within each cluster and speed up the hierarchical clustering process compared with other methods. 

\subsubsection{Effectiveness of Cluster-aware Target Attention}

In the proposed cluster-aware target attention, as compared to TWIN, the main difference lies in using clustered behavior as input and reweighting the attention scores based on cluster size.
We separately removed the cluster size reweighting part from GSU and ESU to assess its impact on performance.
In specific, we replaced the $\boldsymbol{\alpha}'$ with $\boldsymbol{\alpha}$ separately for both GSU and ESU in Eq.~\eqref{eq: alpha'}.
The right part of~\autoref{fig: ablation study} shows the experimental results.
Omitting reweighting leads to performance decline, validating the effectiveness of adjusting attention scores through cluster size.

\subsection{Online Experiments}
\label{sec: online AB}

We conducted an online A/B test to validate the performance of \ourname in our industrial system.
\autoref{tab::abtest} illustrates the relative improvements of \ourname in terms of watch time and diversity of recommended results in three representative scenarios in Kuaishou (Featured-Video Tab, Discovery Tab, and Slide Tab).
Watch Time measures the total amount of time users spend.
Diversity refers to the variety in the model's recommended outcomes, such as the richness in types of videos.
From the results, it is evident that \ourname can better model user interests, leading to improved watch time. Additionally, by modeling longer historical behaviors, \ourname uncovers a more diverse range of user interests, resulting in more varied and diverse recommended results.

\begin{table}[t]
\small
\centering
\caption{Results from online A/B test. 
Both watch time and the diversity of recommended results demonstrated improvements compared with TWIN.
}
\vspace{-10px}
\label{tab::abtest}
\setlength{\tabcolsep}{0.9mm}{\begin{tabular}{lccc}
\toprule
\textbf{Scenarios}
& \textbf{Featured-Video Tab}
& \textbf{Discovery Tab}
& \textbf{Slide Tab}
 \\ 
\midrule
Watch Time   
& +0.672\% & +0.800\% &+0.728\%  \\
Diversity  
& +0.262\% & +0.740\% &+0.005\%\\
\bottomrule            
\end{tabular}
}
\begin{tablenotes}
    \item 0.1\% increase is a significant improvement that brings great business gains in Kuaishou.
\end{tablenotes}
\vspace{-18px}
\end{table}

\section{Conclusion}
In this paper, we propose the \ourname, which effectively extends the maximum length of user history to the life-cycle level, accommodating up to $10^6$ behaviors in Kuaishou.
The offline hierarchical clustering and feature extraction methods compress ultra-long behaviors into shorter clusters, significantly reducing the storage and computational overhead for life-cycle behaviors by 90\%.
The cluster-aware target attention for online inferring captures comprehensive and multi-faceted user interests, leading to more accurate and diverse recommendation results.
Extensive offline and online experiments have demonstrated the effectiveness of \ourname over SOTA baselines.
\ourname has been successfully deployed in Kuaishou, serving the main traffic of around 400 million active users daily.


\bibliographystyle{ACM-Reference-Format}
\balance
\bibliography{main}

\newpage
\appendix
\section{APPENDIX}

\subsection{Summary of Notations}
\label{appendix: notation}
We summarize the important notations used in Section 3 in the following table:
\begin{table}[h]
    \small
    \caption{Important Notations Used in Section 3.}\label{notations}
    \vspace{-15px}
    \begin{center}
    \begin{tabular}{ll|ll|ll}
       \toprule
              $f$ & predictor		&	$\sigma$ &sigmoid 			& $\mathbf x$&feature vector	\\
    $\mathcal D$&dateset			&	$\mathbb R$ & real number set  	&$\hat y$ & predicted CTR \\
    $\mathcal L$&loss& $d_0$& feature dimension		&	$y$& ground truth label\\
    \midrule
    \end{tabular}
    \begin{tabular}{ll|ll}
    $S$ & historical behaviors & $\mathbf{K}$ &  cluster embeddings of $C$ \\
    $C$ & clustered behaviors & $P$ &playing completion ratio of $S$ \\
    $s_j$ & $j$-th behavior in $S$ & $p_j$& playing completion ratio of $s_j$ \\
    $c_i$ & $i$-th cluster in $C$ & $M$& group number \\
    $T$ & length of $S$ & $L_m$& $m$-th item group \\
    $\hat T$ & length of $C$ & $\gamma$ &maximum cluster size \\
    $\mathbf{K}_s$ & item embeddings of $S$ & $\delta$& adaptive cluster number \\
    
    \midrule
    \end{tabular}
    \begin{tabular}{ll|ll}
    $\mathbf{x}_v$ & feature vector of item $v$ & $\mathbf{c}_i$ & feature vector of cluster $c_i$ \\
    $\mathbf{x}_{1:N_1}^{(v)}$ & categorical part of $\mathbf{x}_v$  & $\mathbf{c}_{1:N_1}^{(i)}$ & categorical part of $\mathbf{c}_i$\\
    $\mathbf{x}_{1:N_2}^{(v)}$ & numerical part of $\mathbf{x}_v$ & $\mathbf{c}_{1:N_2}^{(i)}$ & numerical part of $\mathbf{c}_i$  \\
    $\mathbf{k}_{i}$ & embedding vector of $c_i$ & $d$ & embedding dimension \\
    \midrule
    \end{tabular}
    
    \begin{tabular}{ll|ll}
    $\mathbf{q}$ & \footnotesize{target item's inherent embedding} & $\boldsymbol{\beta}$ & \footnotesize{learnable cross feature weight} \\
    $\mathbf{K}_h$ &  inherent feature part of $\mathbf{K}$ & $\mathbf{K}_c$ & cross feature part of $\mathbf{K}$ \\
    $H$ &  inherent feature dimension & $C$ & cross feature dimension \\
    $\boldsymbol{\alpha}$ &  attention weight & $\boldsymbol{\alpha}'$ & adjusted attention weight \\
    $\mathbf{n}$ & cluster size of $C$ & $\mathbb{N}$ & natural number set \\
    
    \midrule
    \end{tabular}
    \begin{tabular}{ll}
    $\qquad\mathbf W^q, \mathbf W^h, \mathbf W^c, \mathbf W^v, \mathbf W^o\qquad$ & $\qquad$linear projection parameters$\quad$  \\
    
    \bottomrule
    \end{tabular}
    \end{center}
    \vspace{-10px}
\end{table}

\subsection{Behavior Feature Splits and Linear Projection}
\label{app: Behavior}
Following TWIN~\cite{TWIN}, we define the feature representations of a length $\hat T$ clustered behavior sequence $[c_1,c_2,...,c_{\hat T}]$ as matrix $\mbf{K}$, where each row denotes the features of one behavior. 
In practice, the linear projection of $\mbf{K}$ in the attention score computation of MHTA is the key computational bottleneck that hinders the application of multi-head target attention (MHTA) on ultra-long user behavior sequences. We thus propose the following to reduce its complexity.

We first split the behavior features matrix $\textbf{K}$ into two parts,
\begin{equation}\label{eq:be_splits}
 \mathbf{K} = [\mathbf{K}_h, \mathbf{K}_c] \in \mathbb{R}^{\hat{T} \times (H+C)},
\end{equation}
We define $\textbf{K}_h \in \mathbb R^{\hat{T} \times H}$ as the \textit{inherent} features of behavior items (e.g. video id,  author, topic, duration) which are 
independent of the specific user/behavior sequence,
and $\textbf{K}_c \in \mathbb R^{\hat T \times C}$ as the user-item cross features (e.g. user click timestamp, user play time, clicked page position, user-video interactions).
This split allows highly efficient computation of the following linear projection $\textbf{K}_h\textbf{W}^h$ and $\textbf{K}_c\textbf{W}^c$.

For the inherent features $\textbf{K}_h$, although the dimension $\textbf{H}$ is large (64 for each id feature), the linear projection is not costly. The inherent features of a specific item are shared across users/behavior sequences. With essential caching strategies, $\textbf{K}_h \textbf{W}^{h}$ could be efficiently ``calculated'' by a look-up and gathering procedure.

For the user-item cross features $\textbf{K}_c$, caching strategies are not applicable because: 
1). Cross features describe the interaction details between a user and a video, thus not shared across users' behavior sequences. 
2). Each user watches a video at most once. Namely, there is no duplicated computation in projecting cross features.
We thus reduce the computational cost by simplifying the linear projection weight. 

  Given $J$ cross features, each with embedding dimension $8$ (since not id features with huge vocabulary size). We have $C = 8J$.
 We simplify the linear projection as follows,
\begin{equation}
\textbf{K}_c \textbf{W}^c \triangleq \left[
\begin{array}{ccc}
\textbf{K}_{c,1} \mathbf w_{1}^c,&...&,\textbf{K}_{c,J}\mathbf w_{J}^c
\end{array}\right]
,
\end{equation}
where $\textbf{K}_{c,j} \in \mathbb R^{\hat T \times 8}$ is a column-wise slice of $\textbf{K}_c$ for the $j$-th cross feature, and $\mathbf w_{j}^c \in \mathbb R^{8}$ is its linear projection weight. Using this simplified projection, we compress each cross feature into one dimension, i.e., $\textbf{K}_c \textbf{W}^c \in \mathbb R^{\hat T\times J}$. Note that
this simplified projection is 
equivalent to restricting $\textbf{W}^c$ to a diagonal block matrix. 

\subsection{Analysis of User Activity Levels}
\label{sec: study on user groups}

\begin{figure}[t]
  \centering
  \includegraphics[width=0.4\textwidth]{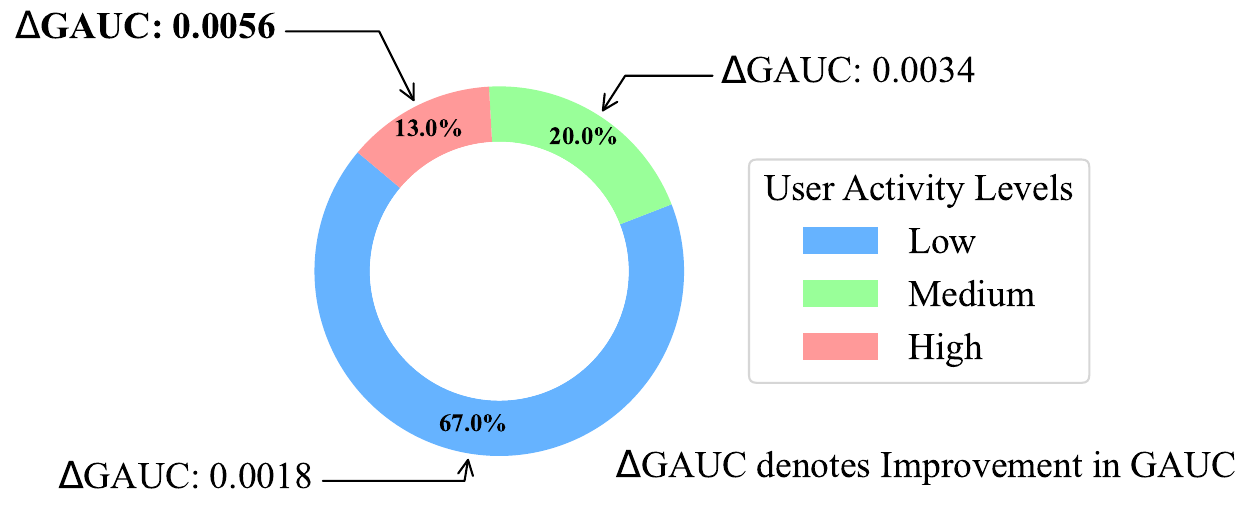}
  \vspace{-10px}
  \caption{GAUC improvement among users of three activity levels with different proportions.
The improvement in GAUC is calculated by subtracting the GAUC of TWIN from the GAUC of \ourname.
\ourname demonstrates more improvements for users with a longer history. 
  } 
  \vspace{-12px}
  \label{fig: GAUC by user groups}
\end{figure}

We postulate that enhancing recommendation model performance can be achieved by extending the length of user history input.
Given that users with different activity levels exhibit varied lengths of historical behavior, the effect of extending long-term interest modeling to the life-cycle level is likely to differ among them.
Consequently, we grouped users by different history lengths and reported the performance improvement across these groups.

We categorized users in the dataset into three groups based on the number of their historical behaviors: Low, Medium, and High.
Additionally, we calculated the GAUC for TWIN and \ourname models across these groups. The improvements in GAUC for TWIN-V2 in different groups and their respective proportions of the total user count are shown in~\autoref{fig: GAUC by user groups}.
It is observable that \ourname achieves performance improvements across all user groups, validating the effectiveness of our approach.
It is also evident that the absolute increase in GAUC is greater in user groups with a higher number of historical behaviors.
This occurs as users with a greater number of historical actions possess a broader spectrum of interests within their life-cycle behaviors, thereby presenting more significant opportunities for enhancement.
Additionally, \ourname also achieves performance improvements in groups of users with shorter histories.
This is due to the incorporation of clustered behavior features in the cluster-aware target attention mechanism. These features represent the aggregated characteristics of all items within the cluster. Consequently, the data fed into GSU and ESU reflects a more comprehensive scope of behaviors compared to TWIN, leading to improved performance.

\end{document}